# Development and Real-World Application of Commercial Motor Vehicle Safety Enforcement Dashboards

**Dhairya Parekh**
Center For Advanced Transportation Technology Laboratory (CATT Lab)
University of Maryland – College Park
College Park, MD 20742 (USA)
Email: dhairyap@umd.edu
(Corresponding Author)

**Mark L. Franz, Ph.D.**
Center For Advanced Transportation Technology Laboratory (CATT Lab)
University of Maryland – College Park
College Park, MD 20742 (USA)
Email: mfranz1@umd.edu

**Sara Zahedian, Ph.D.**
Center For Advanced Transportation Technology Laboratory (CATT Lab)
University of Maryland – College Park
College Park, MD 20742 (USA)
Email: szahedi1@umd.edu

**Narjes Shayesteh**
Center For Advanced Transportation Technology Laboratory (CATT Lab)
University of Maryland – College Park
College Park, MD 20742 (USA)
Email: narjessh@umd.edu

# KEY FINDINGS
- CMV Safety Enforcement Dashboards leverage a data-driven approach to assist with planning and evaluating the effectiveness of enforcement activities instead of relying on anecdotal evidence, local complaints, or standardized FMCSA recommendations
- Development of a data pipeline to integrate diverse data platforms into an intuitive dashboard for safety enforcement planning and evaluation
- The study automated processes to identify critical safety concern areas and collaborated with MSP to plan and evaluate the productivity of a real-world enforcement initiative on I-81 in Hagerstown, MD

# ABSTRACT

Commercial Motor Vehicle (CMV) safety is crucial in traffic management and public safety. CMVs account for numerous traffic incidents, so monitoring CMV safety and safety inspections is essential for ensuring safe and efficient highway movement. This paper presents the development and real-world application of CMV dashboards designed under the guidance of CMV safety enforcement professionals from the Maryland State Police (MSP), the Maryland Department of Transportation - State Highway Administration (MDOT - SHA), and the Federal Motor Carrier Safety Administration (FMCSA) to enable intuitive and efficient analysis of CMV safety performance measures. First, three CMV safety dashboards enable CMV safety professionals to identify sites with a history of safety performance issues. A supplemental dashboard automates the analysis of CMV enforcement initiatives using the same performance measures. These performance measures are based on CMV probe vehicle speeds, inspection/citation data from Truck Weigh and Inspection Stations (TWIS), patrolling enforcement, and Virtual Weigh Stations (VWS). The authors collaborated with MSP to identify a portion of I-81 in Maryland, susceptible to improvement from targeted CMV enforcement. The supplemental enforcement assessment dashboard was employed to evaluate the impact of enforcement, including the post-enforcement halo effect. The results of the post-enforcement evaluation were mixed, indicating a need for more fine-grained citation data.

# KEYWORDS

# INTRODUCTION

The paper introduces a state-of-the-practice dashboard package designed specifically for commercial motor vehicle (CMV) enforcement planning and post-enforcement evaluation. These dashboards aimed to provide valuable insights and tools to improve safety outcomes for CMVs by integrating various data sources, including probe speed data, state police incident data, Federal Motor Carrier Safety Administration (FMCSA) inspection and citation data, and Virtual Weigh Station (VWS) data. The Data and Methods section offers further insights into the data collection and methodology employed in VWS. CMV safety is a critical concern in traffic management and public safety. CMVs account for a significant portion of traffic incidents. Due to their larger size and heavier weight, commercial vehicles are more likely to cause severe injury and higher fatality rates compared to other types of crashes (*1*). In 2021, the National Highway Traffic Safety Administration (NHTSA) reported 61,332 vehicles involved in fatal crashes on U.S. roads. Large trucks account for about 9% of these vehicles (*2*). According to FMCSA, the number of large trucks involved in fatal crashes rose by 18%, from 4,821 to 5,700, from 2020 to 2021. Additionally, the large truck involvement rate in fatal crashes per 100 million vehicle miles travelled increased by 8%, from 1.62 to 1.74 (*3*).

Implementing effective safety measures is crucial, given the substantial impact of CMV-related incidents. Enforcement is one of the most effective strategies for enhancing the safety of all motor vehicles, including CMVs. High Visibility Enforcement (HVE) programs are traffic safety initiatives designed to deter dangerous driving behaviors and promote voluntary compliance with traffic laws. These programs aim to increase the perceived risk of detection by significantly increasing police presence and visibility on roadways. HVE programs have indicated positive impacts on unsafe driving behaviors (*4*). For instance, NHTSA conducted HVE campaigns in Hartford, Connecticut, and Syracuse, New York, targeting distracted driving from handheld cell phone use. The campaigns included dedicated law enforcement, media messages, and before-and-after evaluations. While the Hartford study showed increased awareness of enforcement efforts, self-reported cell phone use while driving remained unchanged.

In contrast, Syracuse experienced a significant reduction in self-reported handheld cell phone use and texting, indicating the effectiveness of HVE enforcement approaches at these locations (*5*). Similarly, a study assessed a high-visibility pedestrian enforcement program in Gainesville, Florida, designed to improve driver compliance with pedestrian right-of-way laws. This program included engineering enhancements, educational efforts, media outreach, enforcement activities, and feedback signs that show the percentage of drivers who yield to pedestrians. The results indicated increased drivers yielding to pedestrians, with higher compliance at natural or unstaged crossings than staged crossings. Staged crossings allow pedestrians to cross the road in two discrete phases, while unstaged crossings are regular pedestrian crossings. While statistical analysis of pedestrian crashes was inconclusive, the program effectively heightened awareness and adherence to pedestrian yielding regulations (*6*).

Other examples of HVE, like "Phone in One Hand, Ticket in the Other," have successfully reduced handheld phone use while driving in California and Delaware (*7*). In West Virginia, an HVE campaign targeting impaired driving and alcohol-related fatalities by increasing sobriety checkpoints led to a notable decline in alcohol-related crash deaths (*8*). Additionally, HVE campaigns in two college towns in Atlanta effectively decreased underage drinking and driving rates, as confirmed by both roadside and web surveys. These findings underscore the success of focused enforcement strategies in enhancing road safety (*9*). Another popular HVE program is the "Click It or Ticket," aimed at increasing seatbelt use (*10*). Initiated in North Carolina in 1993, the program significantly improved seatbelt usage (*11*). Click It or Ticket was later adopted by South Carolina in 2000, followed by all eight southeastern states by 2001 (*12*, *13*). By 2002, ten additional states had adopted the program, with the highest seatbelt usage increases in states fully implementing it (*11*).

In 2004, Washington State adopted the NHTSA's HVE model from the "Click It or Ticket" campaign to address unsafe driving around CMVs through the Ticketing Aggressive Cars and Trucks (TACT) program. The TACT program was an 18-month pilot program modelled after the "Click It or Ticket" campaign, aimed to reduce dangerous driving behaviors around commercial vehicles through enforcement, education, and evaluation (*14*). Factors for selecting study locations included commercial vehicle crash rates, the mean daily traffic, the proportion of commercial vehicles in the traffic, and existing citations for unsafe driving. The TACT Committee also assessed media market availability, potential media spillover, feasibility of aircraft-assisted enforcement, and road conditions. Based on these factors, four high-crash interstate corridors were chosen, with two receiving media messages and increased enforcement, while two served as controls. Results showed significant increases in driver awareness of TACT messages at intervention sites, leading to a 23% to 46% reduction in violation rates and improved driving behaviors (*15*).

Additionally, a TACT program in Indianapolis positively impacted reducing aggressive driving behaviors like speeding, tailgating, and improper lane changes, although these effects were not sustained post-enforcement (*16*). Another study in Alabama evaluated the effectiveness of the TACT program in reducing severe crashes involving passenger and commercial vehicles. Data from crash statistics, officer reports, and video observations revealed that medium and high levels of enforcement significantly reduced crash rates compared to low enforcement. Recommendations include enhancing public awareness and refining evaluation methods (*17*).

Even though many studies examine the impact of enforcement programs, such studies often fail to assess the long-term effectiveness of these initiatives (*4*). Additionally, only a handful of studies describe enforcement planning and provide a framework. Considering the enhanced availability of traffic data and advancements in analytical tools to process large amounts of data, this paper aims to fill this gap by introducing a comprehensive, data-driven set of dashboards designed to plan and assess targeted enforcement. These dashboards include:

**Safety Explorer Dashboards:**
    a. **Speed and Crash View:** Provides network-wide, segment-level speed and crash count information and analytical tools to compare metrics, such as percentage times speeding, aggregated speed data, or crash hotspots between sets of segments.
    b. **Inspection and Citation View:** Utilizes FMCSA inspection and citation data along with VWS data to visualize the temporal variation of inspection and citation counts at the state level and VWS locations. Similar to the "Speed and Crash View," this dashboard also offers the option to compare sets of VWS locations to inform targeted enforcement.
    c. **Detouring View:** Leverages probe trajectory data to detect and visualize detouring routes of commercial vehicles.

**Enforcement Assessment Dashboard:** This dashboard integrates all the datasets to create before-and-after comparison tables that evaluate the effectiveness of enforcement based on various metrics, such as speed, crash counts, and inspection and citation counts.

The first three dashboards are primarily used for planning an enforcement blitz, while the Enforcement Assessment Dashboard focuses on evaluating the effectiveness of enforcement activities. These dashboards were developed under the guidance of CMV enforcement professionals from the Maryland State Police (MSP), Maryland Department of Transportation (MDOT), and FMCSA to ensure the dashboards met the needs and expectations of end users. To show the usefulness of the dashboards, the authors of this paper worked with MSP to plan and evaluate parameters of a real-world enforcement blitz on the stretch of I-81 in Hagerstown, Maryland. Though these dashboards were applied for CMV enforcement, they were designed to ingest and provide insights on any motor vehicle enforcement data.

The remainder of the paper is organized as follows. The next section presents the data and methodology used to develop the dashboards and the dashboard design considerations. The following section presents a real-world application of the dashboards to select the enforcement blitz location on I-81 and evaluate the safety impacts.

## DATA AND METHODS

### Data Processing

This section highlights how the four primary data sources were ingested and processed to deliver a data-driven approach for planning and executing an impactful enforcement initiative.

### Probe Speed Data

This study used preexisting speed data collected from a sample of CMVs on National Highway System (NHS) roadways for the year 2023. The NHS network is divided into directional segments covering approximately 12,000 miles of roadways in Maryland and into approximately 20,000 segments called Traffic Message Channel (TMC) segments. The data are available at 5-minute epochs but were aggregated to an hourly level to demonstrate trends across different peak periods of the day and to facilitate optimized data processing for dashboards.

It was noted that the general traffic stream had a lower percentage of missing probe speed data (38.17%) compared to the specific data for CMVs (53.07%). The dashboard indicated instances of missing data to the end user. Despite these limitations, the CMV probe data played a critical role as it was the only source that could effectively demonstrate changing trends with CMVs, unlike other probe speed data sources that may impute data to fill in the gaps and do not offer vehicle classification differentiation.

For the analysis, speed limits were conflated using the Maryland Department of Transportation—State Highway Administration (MDOT-SHA) data. The probe data's reference speed field was also employed to address data quality concerns. The reference speed on a road segment reflects free-flowing traffic conditions. The reference speed was calculated based on the 95th percentile of observed speeds between 10 p.m. and 5 a.m., assuming minimal congestion during this timeframe.

Additionally, two over-speeding indicators were generated to showcase hourly segment-level trends. The first indicator highlighted instances when speeds exceeded the posted speed limit, while the second indicator identified instances of over-speeding in uncongested traffic flow using the equation:

$$\text{Overspeeding (Uncongested)} = \frac{n(\text{Probe Speed} > \text{Posted Speed Limit})}{n(\text{Probe Speed} > (0.8 * \text{Reference Probe Speed}))} * 100$$

Furthermore, parameters such as 'Historical Average Speed,' which represented the average speed for the roadway segment for the given hour and day of the week, and 'Travel Time,' calculated as the ratio between the length of the roadway segment and the harmonic average probe speed, were also considered in our analysis.

## Crash Data

The police crash reports are submitted to the MSP through the Automated Crash Reporting System (ACRS). Efforts were made to source and conflate crash data to specific roadway segments. This process involved utilizing report numbers to identify whether a CMV was involved in the crash, enabling an independent display of such incidents by injury severity. Information on the location of the collision, including details such as directionality, is derived from multiple columns and processed to associate these incidents with OpenStreetMap (OSM) segments accurately. Once conflated, the data could be seamlessly used with OSM or TMC segment networks, aligning with the dashboard's functionality, where this data was displayed. The conflated data was subsequently merged with the rolled-up hourly probe speed data based on the timestamp from the reports and the segment information.

## Anonymized Trajectory Data

Anonymized trajectory data provides information on commercial vehicle movements. It is typically collected from truck telematics, navigation Global Positioning System (GPS) devices in connected vehicles or mobile devices, enabling the anonymized tracking of movements. However, this data is only available for a portion of the overall traffic stream. To ensure accuracy, a validation analysis verified that the probe trajectory data accurately represented the CMV traffic stream *(19)*.

This dashboard utilized only commercially available, anonymized freight trajectory data for Maryland and matched the waypoints with OSM segments. However, discontinuities in CMV routes within the trajectory data were observed. To address this issue, the Trip Analytics tool from the Regional Integrated Transportation Information System (RITIS) suite *(19)* was employed to fill the trajectory data gaps. This enabled a more precise and comprehensive analysis of detour routes in the study. The anonymized trajectory data from 2022 was utilized to facilitate the analysis of detour route usage for a complete year, thereby mitigating any seasonal fluctuations or route utilization patterns that emerged as a consequence of a road construction project.

Detours taken by CMVs can lead to increased traffic on local roads not designed to handle such vehicles, potentially causing disturbances and property or pavement damage. The Trip Analytics (TA) tool provided a convenient method for studying detoured routes by utilizing a bounding box to filter trips originating or ending outside of that box but with at least one waypoint within it. This method allowed for the study of detoured routes taken for the same origin-destination (OD) pair, specifically to avoid enforcement *(19)*.

Our analysis encompassed all the static enforcement sites in Maryland, including TWISs and VWSs. Route information is obtained from OSM metadata due to its extensive coverage. Meanwhile, average travel times and the number of trips were read from the TA tool output. Notably, the total number of crashes associated with the detouring routes raises safety concerns, emphasizing the need for further investigation and potential mitigation strategies.

## Inspections and Citations Data

The enforcement data, which includes information about the CMV inspections and the citations issued to the vehicles found in violation of any state or federal laws, was accessed through two different sources:
   a. **Federal Motor Carrier Safety Administration (FMCSA)**

The Safety Measurement System (SMS) portal was accessed to obtain the FMCSA enforcement data *(20)*. The data includes information about the number of inspections conducted by MSP, including inspections at TWIS. Moreover, it captures any violations related to the driver or CMV equipment that may have occurred during these inspections. These violations are categorized into the following groups: Vehicle Maintenance, Unsafe Driving, Hours-of-Service Compliance, Driver Fitness, Hazardous Materials Compliance, and Controlled Substances and Alcohol.

If a violation indicates that the driver or the CMV is unfit for the road, it is recorded as an Out-of-Service (OOS) violation, highlighted in the showcased data. Note that this data lacks specific times of day, so the numbers are presented only on a daily basis. Additionally, due to the absence of spatial information, the enforcement numbers cannot be broken down to segment-level data and are only available at the state level.
   b. **Virtual Weigh Stations (VWS)**

VWS comprises well-calibrated sensors that function similarly to stationary enforcement sites like TWIS *(18)*. While identifying the vehicle classification, these automated stations can accurately record various parameters, including vehicle speed, weight, distribution over the axles/load balance, and length. VWS flag violations akin to the FMCSA categories, effectively inspecting and flagging each passing vehicle that does not comply with safe operational rules. However, these stations don't issue automated tickets for violations but are used as a supplementary tool to support an on-site state trooper. The data obtained from these automated stations is accompanied by complete timestamps, thus supporting the identification of hourly trends. Given the widespread coverage of VWS locations across the state, spatial trends can also be addressed, making them valuable sources for comprehensive traffic and enforcement data.

**Dashboard**

Each dashboard was designed under the guidance of CMV safety professionals from FMCSA, MSP, and MDOT. This guidance ensured that the dashboards met the needs of end users.

**Planning - Safety Explorer Dashboards:**

To enhance enforcement officers' ability to pinpoint areas where specific safety metrics were underperforming, the Safety Explorer umbrella encompasses three dashboards: the Speed and Crash View, Inspection and Citation View, and Detouring View. These dashboards, developed using Tableau Desktop, are designed to provide a comprehensive overview of the data. The Speed and Crash View and the Inspection and Citation View feature layers enable temporal and spatial filtering to adjust the resolution as needed. Additionally, the dashboards include options to create a 'Reference Set' and a 'Target Set' for spatial component comparison in the same temporal setting.

    a. **Speed and Crash View:**

In the "Speed and Crash View," shown in Figure 1, safety metrics such as aggregated probe speed, crashes (classified for all vehicle classes or CMVs only), and percentage times speeding (over-speeding indicators) are available. Users start by defining temporal filters such as Year, Month, Week Number, Day of the Week, and Hour of the day, followed by spatial components for the reference and the target set. At least one corridor is required for the reference and target sets, respectively. For example, users can compare I-95N with MD-200W within the defined temporal configuration. The spatial resolution supports narrowing down to TMC segment(s) for comparison by selecting them on the map.

Once all the filters are finalized, the results can be viewed. The chosen safety metric is broken down by temporal charts showing trends at different levels, including annual, monthly, day of the week, and hour of the day. This helps to identify patterns for the selected roadway segments. A roadway segment-level summary table includes information about the posted speed limit, aggregated probe speed, percentage times speeding components, and the minimum and maximum observed probe speeds for the selected temporal and spatial resolution. The table is supplemented with crash data broken down by injury severity and vehicle classification, aiding in identifying roadways that raise safety concerns for over-speeding or are prone to incidents.

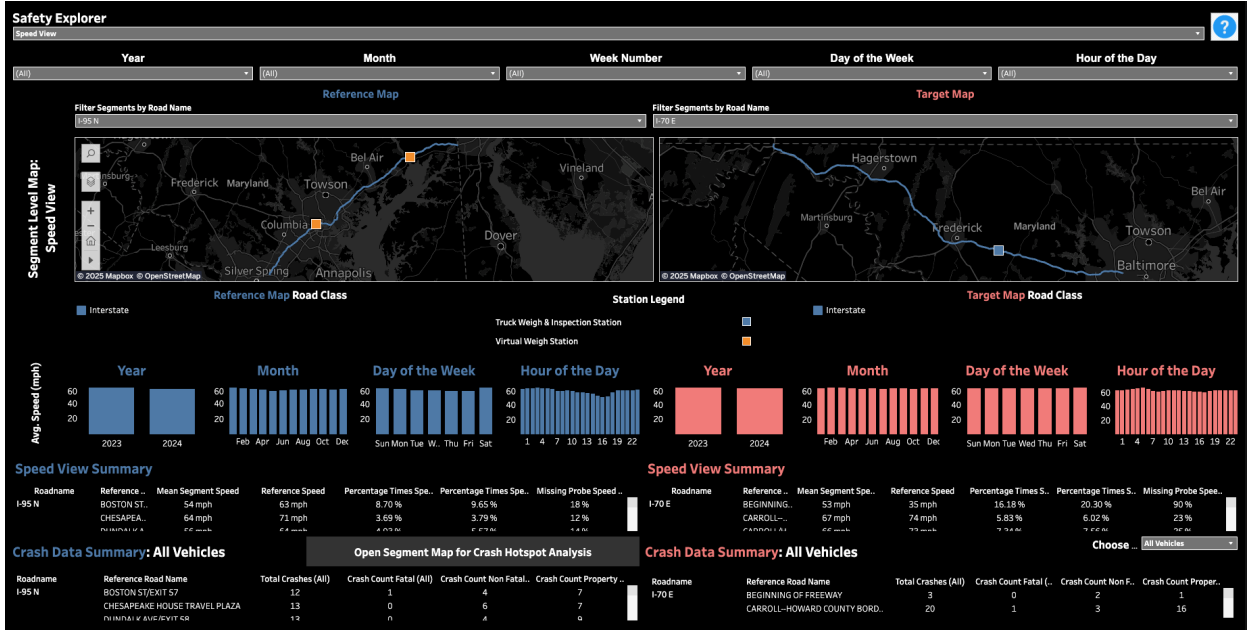

**Figure 1:** Safety Explorer Dashboard: Speed and Crash View

### b. Inspection and Citation View

The following view shares similarities with the "Speed and Crash View" in terms of allowing the selection of temporal and spatial resolution. In this view, users can pick between inspection and citation data from FMCSA or VWS. When opting for the FMCSA data view, a map that displays all TWIS locations is provided solely for reference, since the data does not have spatial information. The FMCSA citation numbers specifically focus on those deemed Out-of-Service (OOS) post-inspection. Figure 2 features a daily trends chart illustrating the correlation between the number of inspections and violations cited throughout the day. These numbers can be presented either in a stacked bar chart format or as a percentage of the total inspections. Furthermore, the breakdown by temporal components, except for the hourly level, is also available.

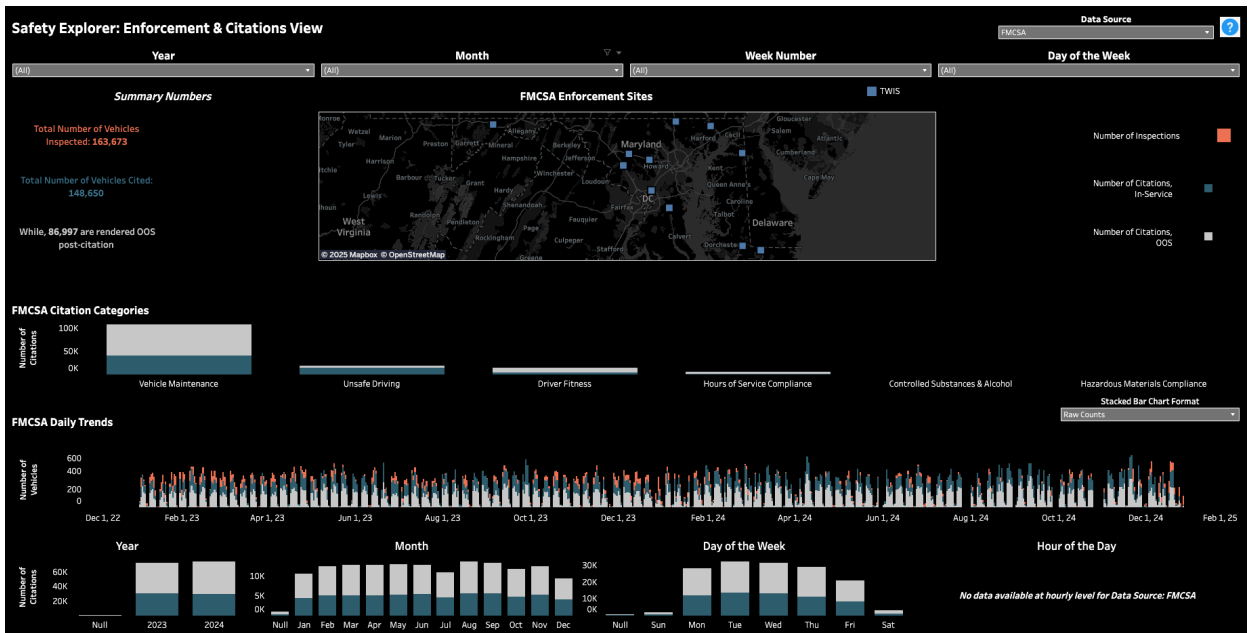

**Figure 2:** Safety Explorer: Inspection and Citation View for FMCSA data

When switching to the VWS view, the temporal filters remain intact but can be adjusted as needed. As shown in Figure 3, users can specify their reference and target VWS sites for comparison within this view. It is recommended that the user choose at least one station for each set or select all VWSs in a specific region and compare them with another area in the state. Users can also filter vehicles based on classifications or their gross weight bins and view data for heavy CMVs in both sets. The VWS citation numbers are categorized and presented visually. A familiar daily trends chart displays the number of vehicles passing through the filtered location(s) in each set and whether they were flagged for violations by the system. Temporal trends, including hourly patterns, conclude the dashboard view.

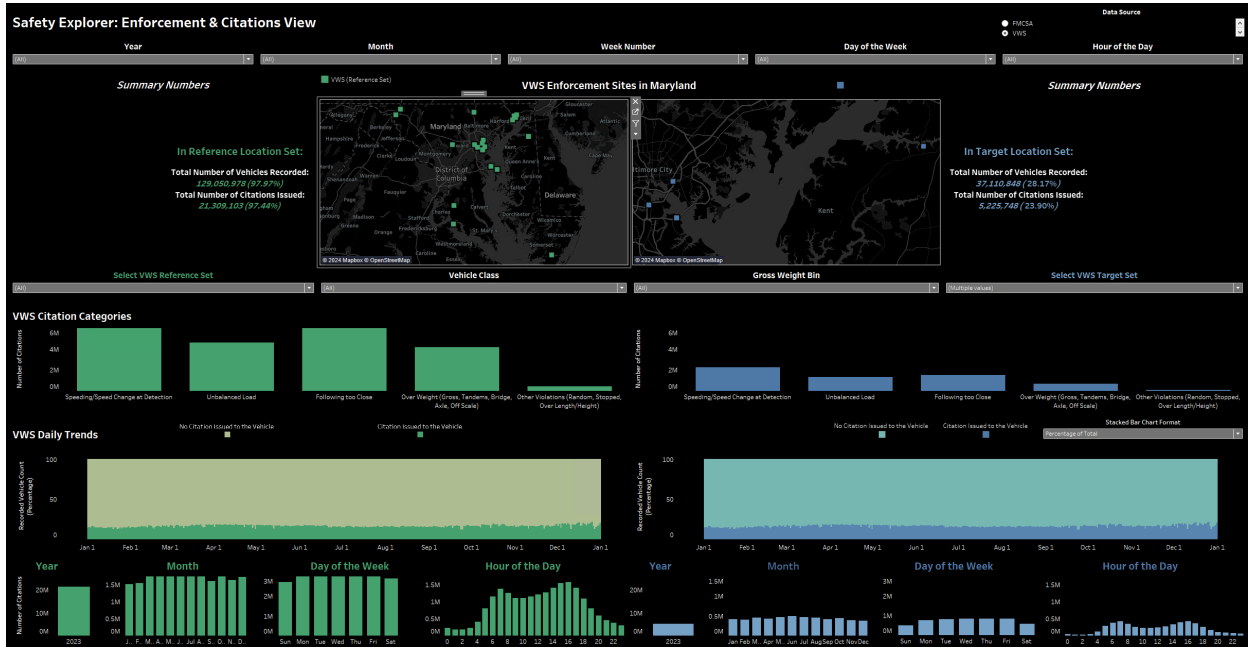

**Figure 3:** Safety Explorer: Inspection and Citation View for VWS data

    c. **Detouring View**

The final safety explorer dashboard specifically focuses on providing valuable information about the alternate routes that drivers commonly use to avoid enforcement. With this dashboard, users can select a stationary enforcement site, either TWIS or VWS, and access detailed route information based on anonymized trajectory data. It is important to note that this dashboard differs from other Safety Explorers as the analysis is static and cannot be adjusted using temporal filters. For instance, when users select an enforcement site, they are presented with results, as shown in Figure 4. These display the potential alternate routes, their corresponding route length, travel time, number of trips, and incidents. This information is crucial in showcasing the practical applications of anonymized trajectory data for effectively planning enforcement areas, maximizing resource utilization, and enhancing the likelihood of initiative success. Note that the anonymized data is historic and privacy-protected, thus it cannot be used for issuing citations to individual drivers. Instead, it simply informs MSP as to which routes have historically been used by drivers so that enforcement officials can determine if additional inspection sites may be needed.

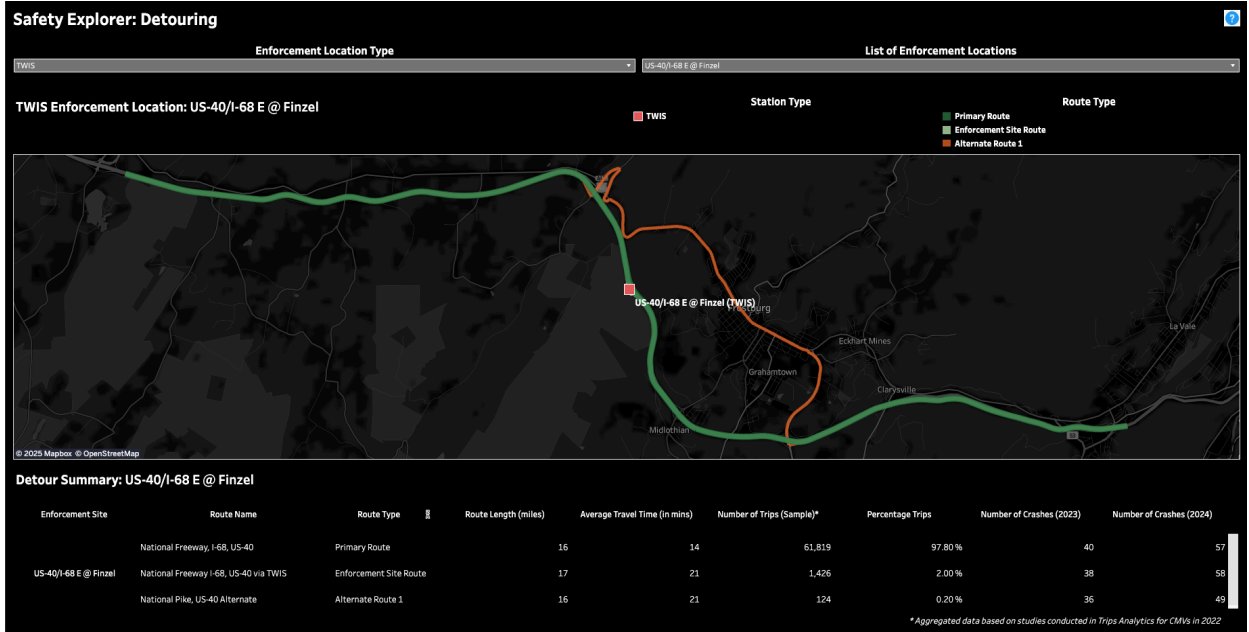

**Figure 4:** Detouring View

## Evaluation - Enforcement Assessment Dashboard

The Safety Explorer dashboards are utilized to plan an enforcement initiative. Following this, the assessment dashboard, Figure 5, can help conduct a before-during-after analysis to assess its outcomes. This dashboard taps into the same data sources used for planning, which include segment-level probe data, FMCSA and VWS enforcement, and citation statistics. What sets it apart is the user's ability to define their before, enforcement, and after periods using temporal filters, select the day of the week or the hour of the day for monitoring, and analyze the data across those periods. Spatially, there are options to filter roadway segments by county, road class, or route name. Segment-level data is available with options to review aggregated probe speed, number of crashes by vehicle classification, and percentage times speeding. Furthermore, filters to narrow down citations from either source are accessible within this view. Additionally, a column displaying percentage change across the before, enforcement, and after periods highlights how the trends with associated metrics have changed, contributing to the decision-making process.

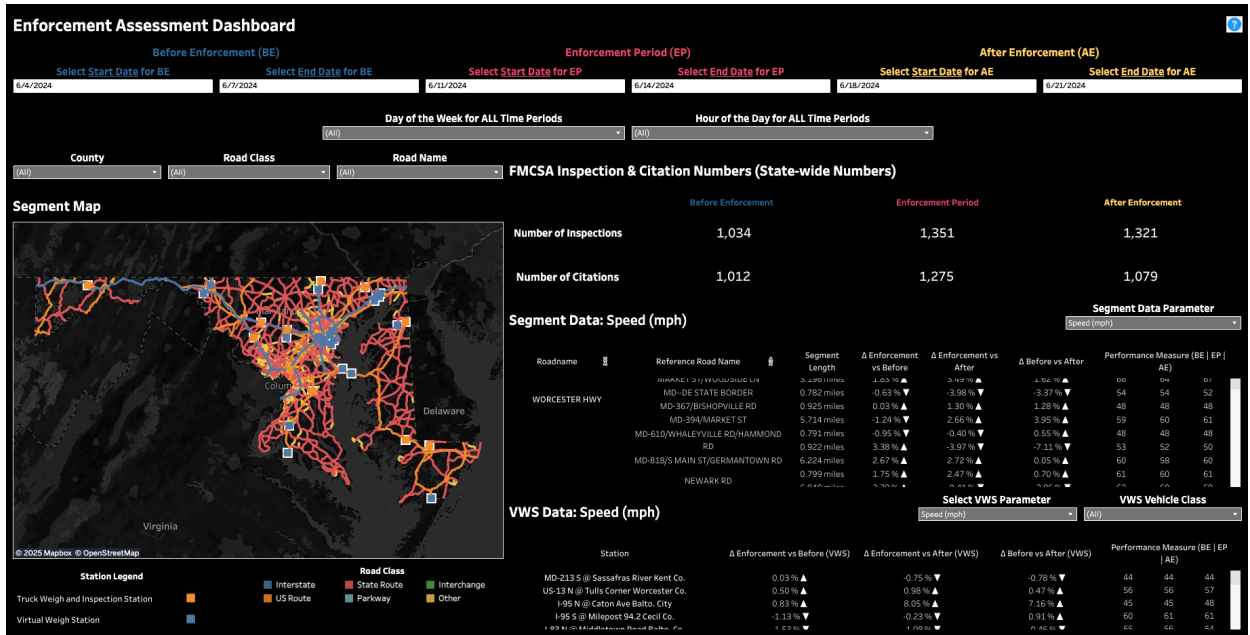

**Figure 5:** Enforcement Assessment Dashboard

## CASE STUDY

In collaboration with MSP, a real-world application for the Safety Explorer dashboards was conducted by identifying potential sites for an enforcement initiative. A data-driven approach was utilized to provide MSP with a detailed recommendation, and the outcomes were demonstrated using the Enforcement Assessment Dashboard.

**Planning the enforcement initiative:**
Candidate sites were selected based on the following criteria, utilizing the Safety Explorer dashboards:
   a. **MSP Crash Data hotspots**
The Speed and Crash View dashboard filtered fatal incidents involving CMVs from 2023. These incidents concentrated around Frederick (US-15), Baltimore County (I-95), and Hagerstown (I-81), so they were added to a list of candidate sites.
   b. **Virtual Weigh Station Citations**
Since VWSs have some spatial information, these citation numbers were preferred over FMCSA. The focus was on speeding and overweight citations since VWSs cited CMVs with those categories the most. When looking at these citation categories, stations in Hagerstown (I-81), Cecil County (I-95), and Baltimore City (I-95 S) reported the highest counts for the aforementioned citation categories.
   c. **Percentage times Speeding**
A pattern was observed where certain time intervals throughout the day had more instances of over-speeding than others. However, due to time constraints imposed by MSP operations, over-speeding was observed only on weekdays between 4 a.m. and 5 p.m. Segment-level probe speed data helped identify Hagerstown (I-81), Howard County (MD-32), and the Eastern Shore of Maryland (US-50/US-301) as the predominant sites where the issue occurred.

**Final case study parameters:**
Hagerstown (I-81) in Washington County was chosen for the planned enforcement site because it was the only site that met all the determined criteria. The site's proximity to state lines with Pennsylvania and West Virginia, different speed limits across the states, and a VWS in both travel directions made it an ideal location for this study. Detour routes determined from the detouring view provided guidance on commonly used enforcement avoidance routes. This information was used to make recommendations for supplemental enforcement using patrol vehicles. A timeline recommendation for the before, enforcement, and after period was established by noting when the VWSs issued the most citations. As a result, the enforcement took place between 6/11/24 (Tuesday) and 6/14/24 (Friday), from 6 am to 2 pm. The previous and the following week, Tuesday-Friday, were selected as the before (6/4/24 - 6/7/24) and after (6/18/24 - 6/21/24) enforcement periods for analysis, respectively. To assess the impact of the enforcement, a special arrangement was made with MSP to track the FMCSA-reported inspection and citation numbers at this location, as the publicly available data lacks location information.

**Evaluating the I-81 enforcement initiative:**
After completing the on-the-ground enforcement activities, data from MSP was used to demonstrate the enforcement initiative's results through the Enforcement Assessment dashboard (Figure 6). The citation numbers from FMCSA were documented for both the I-81 initiative and statewide numbers (Table 1). Note that the MSP/FMCSA citation data did not include directionality information. Unfortunately, MSP did not provide any data for inspections during the after-period. Thus, the evaluation of the enforcement halo effect was not possible. As shown in Table 1, the number of I-81 inspections increased by 21% between the before and during enforcement periods. The trend was also present in the state-wide inspections, where the number of inspections increased by 31%. However, the citations per inspection on I-81 increased from 0.77 in the before-enforcement period to 2.04 in the enforcement period. The state-wide metric had an opposite trend, with the citations per inspection decreasing from 0.98 in the before-enforcement period to 0.94 during enforcement to 0.82 in the after-enforcement period. While inspections increased on I-81 and across the state, the enforcement initiative on I-81 discovered more citations per inspection relative to the before-enforcement period.

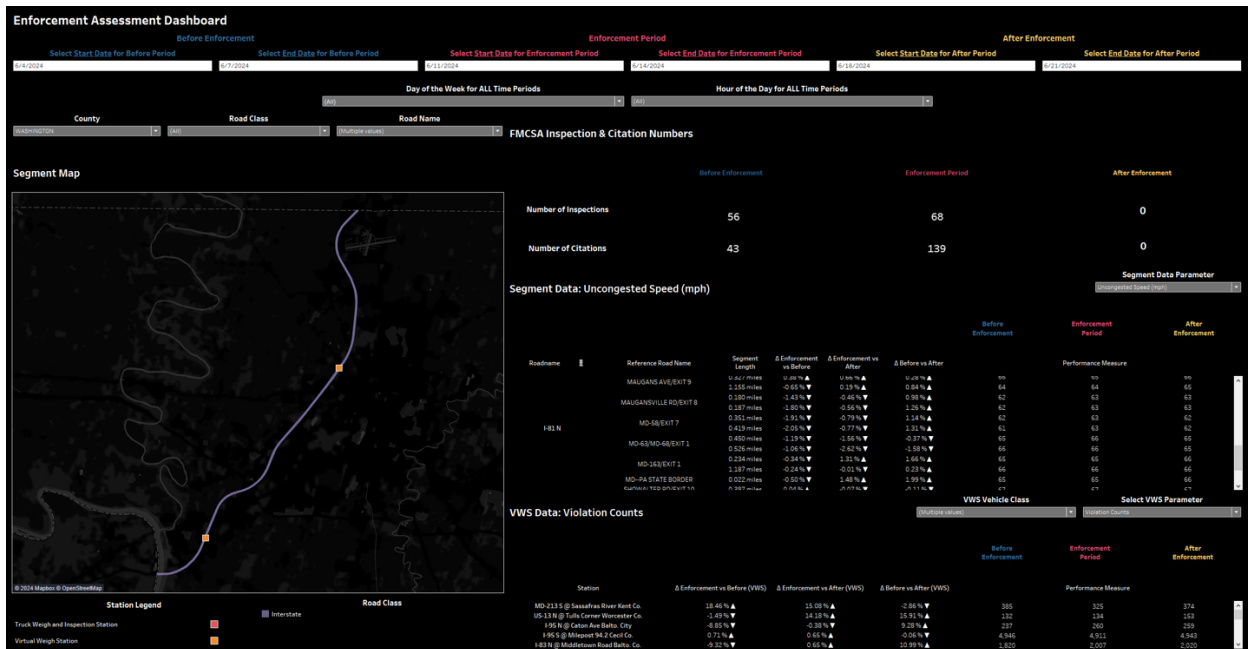

**Figure 6:** The result of the I-81 initiative showcased using the Enforcement Assessment Dashboard

Next, citations from both VWS stations on I-81 were evaluated. As presented in Table 1, the VWS on I-81N experienced a decrease of 2.2% in citations between the before and during enforcement periods. However, citations increased by 3.1% between the enforcement and post-enforcement periods. These findings suggest that the enforcement initiative, which included the presence of MSP patrol vehicles, improved safety on I-81N by reducing violations. However, I-81N did not benefit from the enforcement halo effect. One potential explanation for this result could be that since the stretch of I-81 is a long-distance travel corridor, it may not experience as much daily traffic from the same CMVs. Consequently, a more local route may be more susceptible to the halo effect.

Interestingly, I-81S experienced a monotonic increasing trend across the analysis periods. This result suggests that the enforcement initiative on I-81S was ineffective in reducing the number of citations. It is worth noting that the details on directionality in the MSP citation data were not provided. Thus, a possible explanation for the unexpected result on I-81 S is that enforcement was more rigorous in the northbound direction.

**Table 1:** Overview of FMCSA Inspection/Citation Numbers for the Initiative and State-wide Numbers with CMV VWS Violation Counts tracked from I-81 stations

|   |   |   | Before Enforcement | Enforcement Period | After Enforcement | Before vs Enforcement (% change) | Enforcement vs After (% change) | Before vs After (% change) |
|---|---|---|---|---|---|---|---|---|
| FMCSA | Number of Inspections | I-81 Initiative | 56 | 68 | 0 | 21.4 | NA | NA |
| | | State-wide | 1036 | 1353 | 1320 | 30.6 | -2.4 | 27.4 |
| | Number of Citations | I-81 Initiative | 43 | 139 | 0 | 223.3 | NA | NA |
| | | State-wide | 1012 | 1274 | 1076 | 25.9 | -15.5 | 6.3 |
| | Citations per Inspection | I-81 Initiative | 0.77 | 2.04 | NA | - | - | - |
| | | State-wide | 0.98 | 0.94 | 0.82 | - | - | - |
| VWS | Number of Citations | I-81 N @ Milepost 1.8 Wash Co. | 6274 | 6137 | 6329 | -2.2 | 3.1 | 0.9 |
| | | I-81 S @ Milepost 7.6 Wash Co. | 5700 | 5983 | 6017 | 5.0 | 0.6 | 5.6 |

Lastly, the CMV probe vehicle speeds from the probe-based speed dataset were analyzed. Table 2 summarizes this analysis, including average speeds and the percentage times speeding during uncongested conditions. In comparing the average CMV probe speed, I-81N experienced a 3.2% increase in speeds during the enforcement period relative to the before period. This observation suggests that the enforcement initiative did not improve safety regarding CMV speeds. There was no change in average speeds between the enforcement and after-enforcement periods. This finding indicates that no enforcement halo effect was present. I-81N experienced an increase in the percent time speeding between the before and during enforcement periods. However, the after-enforcement period experienced a return to the before-enforcement percentage times speeding. I-81S experienced a monotonic increase in average speed throughout the analysis period, suggesting that the enforcement did not improve safety performance in terms of average speed. I-81S percent time speeding was consistent at 22% for all three periods, suggesting that enforcement had no impact, and the halo effect was absent.

**Table 2:** Overview of Segment-level CMV Probe Data (Monitored between 6 am - 2 pm)

|   |   | Before Enforcement | Enforcement Period | After Enforcement | Before vs Enforcement (% change) | Enforcement vs After (% change) | Before vs After (% change) |
|---|---|---|---|---|---|---|---|
| Avg Probe Speed (Uncongested) (mph) | I-81 N | 63 | 65 | 65 | 3.2 | 0.0 | -3.2 |
| | I-81 S | 62 | 63 | 64 | 1.6 | 1.6 | -3.2 |
| Percentage times Speeding (%) | I-81 N | 19% | 21% | 19% | -2.0 | -2.0 | 0.0 |
| | I-81 S | 22% | 22% | 22% | 0.0 | 0.0 | 0.0 |

Note that the MSP citation data did not include directionality. However, the data suggest that the initiative was effective overall in increasing the number of inspections, citations, and citation rates per inspection.

The results of the analysis of the enforcement initiative were mixed for I-81N. The study of I-81N VWS data showed a decrease in citations during the enforcement period, which supports the argument that enforcement improved safety. However, in analyzing average CMV probe speeds and percent time speeding, the results showed increasing average speeds during the enforcement period. The enforcement halo effect was absent in any of the metrics on I-81N.

The results on I-81S indicated that enforcement did not improve safety, as VWS citations and average CMV probe vehicle speeds monotonically increased over the analysis periods. The percentage times speeding was consistent across each analysis period.

## CONCLUSIONS

The paper presents the details of developing a suite of automated analysis dashboards to support CMV safety enforcement planning and evaluation. The dashboards leverage data commonly collected by DOTs and their enforcement partners to provide data-driven decision guidance for CMV enforcement activities. All dashboards were developed under the guidance of CMV safety experts. The Safety Explorer Dashboards allow end users to investigate the historical patterns of key safety performance measures, including CMV speeds, crashes, inspection and citations, and detour/enforcement avoidance route usage. These dashboards were designed to enable CMV enforcement decision-makers to discover a short list of potential enforcement initiative locations with a few clicks of a mouse. Once the enforcement initiative is completed, the Enforcement Assessment Dashboard allows end users to assess the impacts of enforcement efforts on the same performance measures used to select the enforcement location. While these dashboards were designed for CMV enforcement planning and evaluation, they can easily be adapted to ingest any data on motor vehicle enforcement.

To demonstrate the dashboards' application, the team collaborated with MSP to apply them to a real-world enforcement initiative. The dashboards were used to identify a candidate enforcement location and evaluate the safety impacts of the enforcement activities. The analysis of MSP/FMCSA, inspection, and citation data suggested that the enforcement initiative improved safety with increased citations and citation rates during the enforcement period. The VWS indicated that the enforcement initiative was effective in terms of decreasing citations during the enforcement period for the northbound direction. However, the southbound direction did not experience these safety benefits. Lastly, the evaluation of CMV vehicle probe average speeds and percent time speeding suggested neither direction of I-81 experienced safety improvement.

These mixed results suggest the need for more investigation into the enforcement effort, such as the directionality of non-VWS inspection and citation data. Additional next steps include outreach to other CMV safety enforcement agencies to explore further collaborations. Such opportunities will also enable more profound insights into state-of-practice for CMV enforcement planning and evaluation, including data requirements and associated performance measures.

## DATA AVAILABILITY STATEMENT

The dashboard's development incorporates a blend of proprietary data and publicly accessible platforms. Probe speed data is sourced from a commercially available speed data collected from a sample of CMVs on the National Highway System. Anonymized trajectory data is proprietary from an industry partner, while FMCSA inspections and citation data are obtained through their SMS portal. The VWS data is made available via MDOT, and the crash dataset originates from MSP's ACRS portal.

## ACKNOWLEDGMENTS

The authors express their sincere gratitude to the Federal Motor Carrier Safety Administration for their guidance in the successful execution of this project. Additionally, they extend their appreciation to the Maryland Department of Transportation and the Maryland State Police for their support and assistance in undertaking the activities associated with this case study.

## AUTHOR CONTRIBUTIONS

The authors confirm their contribution to the paper as follows:
- Study conception and design: M.L. Franz, and D. Parekh
- Data acquisition/collection: D. Parekh, and S. Zahedian
- Analysis and interpretation of results: M.L. Franz, D. Parekh, and S. Zahedian
- Draft manuscript preparation: D. Parekh, M.L. Franz, S. Zahedian, and N. Shayesteh

All authors have read and agreed to the published version of the manuscript.

## FUNDING

The Federal Motor Carrier Safety Administration partially funded this research. Agreement Number: 69A3602241001MHP0MD. Period of performance: July 1, 2022 – September 30, 2024

## CONFLICTS OF INTEREST

The authors declare that there are no conflicts of interest.

13. Solomon, M. G. (Mark G., R. G. Ulmer, and D. F. Preusser. Evaluation of Click It or Ticket Model Programs. I. Preusser Research Group, ed.

14. Nerup, P., P. Salzberg, J. Van Dyk, L. Porter, R. D. Blomberg, F. D. Thomas, and L. A. Cosgrove. Ticketing Aggressive Cars and Trucks in Washington State: High Visibility Enforcement Applied to Share the Road Safely.

15. Thomas, F. D., R. D. Blomberg, R. C. Peck, L. A. Cosgrove, and P. M. Salzberg. Evaluation of a High Visibility Enforcement Project Focused on Passenger Vehicles Interacting with Commercial Vehicles. Journal of Safety Research, Vol. 39, No. 5, 2008, pp. 459–468. https://doi.org/10.1016/j.jsr.2008.07.004

16. Tarko, A. P., P. C. Anastasopoulos, and A. M. P. Zuriaga. Can Education and Enforcement Affect Behavior of Car and Truck Drivers on Urban Freeways. 2011.

17. Dye, D. C. Evaluation of the Alabama Ticketing Aggressive Cars and Trucks Selective Enforcement Program. The University of Alabama, 2016.

18. Franz, M., Zahedian, S., Parekh, D., Emtenan, T., & Jordan, G. (2024). Exploring Commercial Vehicle Detouring Patterns through the Application of Probe Trajectory Data. ArXiv. /abs/2407.17319

19. Center for Advanced Transportation Technology Laboratory (CATT Lab). Trip Analytics within the Regional Integrated Transportation Information System (RITIS). https://trips.ritis.org Accessed July 26, 2023.

20. Federal Motor Carrier Safety Administration, Safety Measurement System. https://ai.fmcsa.dot.gov/SMS/Tools/Index.aspx. Accessed February 2, 2024
15